\documentclass[superscriptaddress,aps,letterpaper,10pt,twocolumn,floats,showpacs,amsmath,amsfonts,amssymb,pre]{revtex4-2}

\usepackage{graphicx}% Include figure files
\usepackage[colorlinks=true,citecolor=blue]{hyperref}
\usepackage[caption=false]{subfig}
\usepackage{pstricks}
\usepackage{pst-node}
\usepackage{amsmath}
\usepackage{amsbsy}
\usepackage{verbatim}
\usepackage{dsfont}
\usepackage[T1]{fontenc}
\begin{document}

%\preprint{APS/123-QED}

\title{Ensemble dependence of information-theoretic contributions to the entropy production}
%\title{Ensemble equivalence and nonequivalence in information-theoretic formulation of entropy production}
% Force line breaks with \\

\author{Krzysztof Ptaszy\'{n}ski}
\affiliation{Complex Systems and Statistical Mechanics, Department of Physics and Materials Science, University of Luxembourg, L-1511 Luxembourg, Luxembourg}
\affiliation{Institute of Molecular Physics, Polish Academy of Sciences, Mariana Smoluchowskiego 17, 60-179 Pozna\'{n}, Poland}
\email{krzysztof.ptaszynski@uni.lu}
\author{Massimiliano Esposito}
\affiliation{Complex Systems and Statistical Mechanics, Department of Physics and Materials Science, University of Luxembourg, L-1511 Luxembourg, Luxembourg}

\date{\today}% It is always \today, today,

\begin{abstract}
The entropy production of an open system coupled to a reservoir initialized in a canonical state can be expressed as a sum of two microscopic information-theoretic contributions: the system-bath mutual information and the relative entropy measuring the displacement of the environment from equilibrium. We investigate whether this result can be generalized to situations where the reservoir is initialized in a microcanonical or in a certain pure state (e.g., an eigenstate of a nonintegrable system), such that the reduced dynamics and thermodynamics of the system are the same as for the thermal bath. We show that while in such a case the entropy production can still be expressed as a sum of the mutual information between the system and the bath and a properly redefined displacement term, the relative weight of those contributions depends on the initial state of the reservoir. In other words, different statistical ensembles for the environment predicting the same reduced dynamics for the system give rise to the same total entropy production but to different information-theoretic contributions to the entropy production.
\end{abstract}

\maketitle

One of the main goals of statistical physics is to rationalize how time-reversal symmetric microscopic laws of classical or quantum mechanics give rise to thermodynamic irreversibility described by the second law of thermodynamics. Recent decades brought much progress in this area, presenting several complementary explanations of the emergence of irreversibility in both closed~\cite{gogolin2016, polkovnikov2011, eisert2015, alessio2016, mori2018} and open~\cite{spohn1978, andrieux2009, esposito2009, jarzynski2011, campisi2011, campisi2011b, strasberg2019, rivas2020, strasberg2021, landi2021, horodecki2013, brandao2013, brandao2015} quantum systems. Among others, the information-theoretic framework proposed in Ref.~\cite{esposito2010} provided a microscopic basis for the nonnegativity of the entropy production -- a key quantity characterizing the irreversibility of thermodynamics processes. This approach is applicable to a generic open quantum system described by the Hamiltonian
\begin{align}
	H=H_S+H_B+H_I,
\end{align}
where $H_S$, $H_B$ and $H_I$ are Hamiltonians of the system, bath, and the interaction between them, respectively. The joint state of the system and the bath $\rho_{SB}$ is assumed to undergo a unitary evolution $i\dot{\rho}_{SB}= [H,\rho_{SB}]$ starting from the initially factorized state $\rho_{SB}(0)=\rho_{S}(0) \otimes \rho_B^\text{th}$, where $\rho_{S}(0)$ is an arbitrary initial state of the system, and $\rho_B^\text{th}=\exp(-\beta H_B)/Z_B$ is the canonical Gibbs state of the environment, with $\beta$ being the inverse temperature of the reservoir, and $Z_B =\text{Tr} \exp(-\beta H_B)$ being the partition function (here and from hereon we take $\hbar=k_B=1$). The entropy production within the time interval $[0,t]$ is defined as
\begin{align} \label{entrprod}
	\sigma = \Delta S_S - \beta Q
\end{align}
where $\Delta S_S=S_S(t)-S_S(0)$ is the change of the von Neumann entropy of the system $S_S=-\text{Tr} (\rho_S \ln \rho_S)$ and $Q={-\text{Tr} \{H_B [\rho_B(t)-\rho_B(0)] \}}$ [with $\rho_B(0)=\rho_B^\text{th}$ for the initial thermal state] is the heat extracted from the environment, defined as the change of the bath energy with a minus sign; the formalism can be easily generalized to the grand canonical ensemble by properly accounting for the chemical work. It was shown that the entropy production can be expressed as a sum of two nonnegative information-theoretic constituents:
\begin{align} \label{entrprodinf}
	\sigma = I_{SB} + D[\rho_B(t)||\rho_B^\text{th}] \geq 0,
\end{align}
where $I_{SB}=S_S(t)+S_B(t)-S_{SB}(t)$ is the quantum mutual information between the system and the bath and $D[\rho_B(t)||\rho_B^\text{th}]=\text{Tr} \{ \rho_B(t) [\ln \rho_B(t)-\ln \rho_B^\text{th}] \}$ is the quantum relative entropy that measures the displacement of the environment from equilibrium. According to information theory, the terms $I_{SB}$ and $D[\rho_B(t)||\rho_B^\text{th}]$ are nonnegative, which provides a microscopic basis for the second law of thermodynamics (see Ref.~\cite{strasberg2021b} for an even tighter bound with finite-size corrections).

As further discussed in Ref.~\cite{ptaszynski2019}, a particularly elegant interpretation of the entropy production is provided by assuming that the environment is composed of $K$ independent degrees of freedom $k$ (later referred to as modes), such that $H_B=\sum_{k=1}^K H_k$. Then Eq.~\eqref{entrprodinf} can be rewritten as
\begin{align} \label{entrprodmod}
	\sigma = I_{SB}+I_\text{env}+ D_\text{env} = I_\text{tot} + D_\text{env}\geq 0,
\end{align}
where $I_\text{env}=\sum_k S_k(t)-S_B(t)$ is the mutual information between the modes of the environment, $I_\text{tot} = I_{SB}+I_\text{env}$ is the total correlation, i.e., a sum of system-bath and intraenvironment correlations, and the term $D_\text{env} =\sum_k D[\rho_k(t)||\rho_k^\text{th}]$ measures the displacement of the modes of environment from equilibrium. For $K \rightarrow \infty$ the contribution $D_\text{env}$ usually tends to be negligible, since each mode is only slightly perturbed from equilibrium (though deviations from this behavior are possible when only a small portion of the environment is resonantly excited~\cite{colla2021}), and thus the entropy production can be related to the generation of multipartite correlations between the system and the modes of environment $I_\text{tot}$. 

In deriving Eq.~\eqref{entrprodinf} one assumes that the initial state of the environment is the canonical Gibbs state. However, it has been shown that certain non-thermal initial states of the bath may lead (under certain conditions) to the same reduced dynamics and thermodynamics of the system as the thermal state; later, such property will be referred to as the \textit{dynamical equivalence} to the canonical state. A first example studied in the literature was the microcanonical state~\cite{esposito2003, esposito2007, riera2021, heveling2022}, namely, an equally-weighted mixture of energy eigenstates of the bath with energies within the interval $[E-\delta,E+\delta]$, where $2 \delta$ is the width of the microcanonical shell. The dynamical equivalence is there a consequence of a well-known principle of equilibrium \textit{ensemble equivalence}~\cite{gibbs1902}, which states that in the thermodynamic limit the microcanonical and canonical states are equivalent with respect to their thermodynamic properties and expected values of observables (see Ref.~\cite{touchette2015} for a contemporary formulation of this concept). While the dynamical equivalence can be formally proven~\cite{esposito2007, riera2021, heveling2022}, here we will present only a qualitative justification. Usually, the system does not interact in a uniform way with the whole environment, but rather is more strongly coupled to some of its (possibly small) parts; for example, a system coupled to a bath of harmonic oscillators (the Caldeira-Leggett model~\cite{caldeira1983}) will most strongly couple to those oscillators whose resonant frequencies are close to transition frequencies of the system. At the same time, as implied by the principle of ensemble equivalence, when the whole environment is initialized in the microcanonical state, a reduced state of its small part (effectively coupled to the system) is the canonical state. As a consequence, the system evolves as if it was coupled to the thermal bath.

Furthermore, there exist also several types of pure states which reproduce equilibrium properties of the thermal state, and thus may be expected to be also dynamically equivalent. First, an equivalence of static observables can be observed even for single eigenstates of nonintegrable systems obeying the eigenstate thermalization hypothesis (ETH)~\cite{deutsch1991, srednicki1994, deutsch2018}; in such a case the dynamical equivalence, namely, the applicability of the second law of thermodynamics and the nonequilibrium fluctuation theorem, has been recently demonstrated~\cite{iyoda2017, iyoda2021, heveling2022}. Other examples are typical superpositions of states from the microcanonical shell~\cite{tasaki1998, goldstein2006, camalet2008, reimann2022}, thermofield double states~\cite{israel1976} (purifications of a thermal state in a doubled Hilbert space), and so-called thermal pure states~\cite{sugiura2013, endo2018, heitmann2020} (coherent superpositions of energy eigenstates with populations obeying the Boltzmann distribution). As a matter of fact, as shown by Popescu \textit{et al.}~\cite{popescu2006}, almost every pure state of the environment leads to relaxation of the system to the canonical state. 

This raises the question of whether the information-theoretic formulation of the entropy production can be generalized to a generic initial state of the bath dynamically equivalent to the thermal state. In this Letter we note that for arbitrary initial state of the bath (initially uncorrelated with the system) Eq.~\eqref{entrprodinf} can be generalized to a form
\begin{align} \label{entrprodinfgen}
	\sigma = I_{SB}+\Delta D(\rho_B||\rho_B^\text{th}),
\end{align}
where $\Delta D(\rho_B||\rho_B^\text{th}) = D [\rho_B(t)||\rho_B^\text{th}]-D\left[\rho_B(0)||\rho_B^\text{th} \right]$ measures the change of displacement of the bath state from equilibrium. This can be easily derived as follows. First, by expanding $D [\rho_B(t)||\rho_B^\text{th}]=\text{Tr} [\rho_B(t) \ln \rho_B(t)]-\text{Tr} [\rho_B(t) \ln \rho_B^\text{th}]=-S_B(t)+\ln Z_B + {\beta \text{Tr} [\rho_B(t) H_B]}$, one gets $\Delta D(\rho_B||\rho_B^\text{th})=-\Delta S_B-\beta Q$, where $\Delta S_B=S_B(t)-S_B(0)$, and the heat $Q$ is defined below Eq.~\eqref{entrprod}. Second, one uses the assumption that the system and the bath are initially uncorrelated, which implies vanishing of the initial mutual information: $I_{SB}(0)=S_S(0)+S_{B}(0)-S_{SB}(0)=0$; as the unitary dynamics conserves the joint von Neumann entropy of the system and the bath $[S_{SB}(t)=S_{SB}(0)]$ this implies $I_{SB}=S_S(t)+S_{B}(t)-S_{SB}(t)-[S_S(0)+S_{B}(0)-S_{SB}(0)]=\Delta S_S+\Delta S_B$. Inserting $\Delta D(\rho_B||\rho_B^\text{th})=-\Delta S_B-\beta Q$ and $I_{SB}=\Delta S_S+\Delta S_B$ into the right hand side of Eq.~\eqref{entrprodinfgen} one gets $\Delta S_S-\beta Q$, which is the entropy production [Eq.~\eqref{entrprod}].

It can be now noted that in general $\Delta D(\rho_B||\rho_B^\text{th})$ is not necessarily nonnegative, and therefore Eq.~\eqref{entrprodinfgen} does not provide a basis for the second law of thermodynamics for non-thermal initial states of the bath; nevertheless, it still enables one to express the entropy production in terms of microscopic, information-theoretic contributions, while its nonnegativity can be provided by the dynamical equivalence with the thermal state. However, we will show that while the relation~\eqref{entrprodinfgen} always holds, the relative weight of the terms $I_{SB}$ and $\Delta D(\rho_B||\rho_B^\text{th})$ may depend on the initial state of the bath; therefore, dynamically equivalent states are only partially equivalent from the perspective of information-theoretic formulation of the entropy production.

For environments composed of independent modes in an arbitrary initial state, Eq.~\eqref{entrprodmod} can be generalized as
\begin{align} \label{entrprodmodgen}
	\sigma =I_{SB}+\Delta I_\text{env}+ \Delta D_\text{env} = \Delta I_\text{tot} + \Delta D_\text{env}.
\end{align}
This can be derived as follows: first, in analogy to the derivation below Eq.~\eqref{entrprodinfgen}, one gets $D_\text{env} =\sum_k D[\rho_k(t)||\rho_k^\text{th}]=-\sum_k \Delta S_k-\beta Q$, where $\Delta S_k=S_k(t)-S_k(0)$; then using $\Delta I_\text{env}=\sum_k \Delta S_k-\Delta S_B$ and $I_{SB}=\Delta S_S+\Delta S_B$, one finds that the right hand side of Eq.~\eqref{entrprodmodgen} is equal to $\sigma=\Delta S_S-\beta Q$. It may be now argued that for initial states of the bath dynamically equivalent to the thermal state, the initial states of the modes are thermal (in the thermodynamic limit of large $K$), and that their reduced states evolve in the same way. As a consequence, the contributions $\Delta D_\text{env}$ and $\Delta I_\text{tot}$ -- which depend on the local states of the modes rather than the total state of the bath $\rho_B$ -- should also be the same. This will be demonstrated later by numerical simulations.

\textit{Example 1: Random matrix Hamiltonian}. Let us now investigate, for two exemplary cases, how  the behavior of information-theoretic constituents of the entropy production depends on the initial state of the reservoir. To this goal, we perform simulations of the unitary dynamics of the system-bath ensemble for a system coupled to a finite environment. First, we will consider a nonintegrable system obeying the eigenstate thermalization hypothesis, defined by means of random matrices; similar setups have been previously investigated in Refs.~\cite{esposito2003, esposito2010}. The Hamiltonian of the system is defined as $H_S=\gamma \sigma_z/2$, $H_B=X_B/\sqrt{8N}$, and $H_I=\lambda \sigma_x \otimes X_I/\sqrt{8N}$, where $X_i$ ($i \in \{B,I\}$) is a Gaussian orthogonal random matrix of size $N$ with variance of the diagonal elements equal to 1, and $\sigma_i$ ($i \in \{x,y,z\}$) are Pauli matrices. As initial states of the bath we took the canonical state, two microcanonical states with different widths of the microcanonical shell $2\delta$, and a single eigenstate of $H_B$ with energy closest to the average energy of the thermal state $\langle E_B^\text{th} \rangle=\text{Tr} (H_B \rho_B^\text{th})$; the microcanonical state is defined as a mixture of eigenstates of $H_B$ with energies $E_i \in [\langle E_B^\text{th} \rangle-\delta,\langle E_B^\text{th} \rangle+\delta]$. In our numerical simulations the random matrices $X_B$ and $X_I$ were generated using the function RandomVariate[GaussianOrthogonalMatrixDistribution[N]] in Mathematica. The joint system-bath state was propagated iteratively as $\rho_{SB}(t+\Delta t)=e^{- iH \Delta t} \rho_{SB}(t) e^{iH \Delta t}$; we used the time step $\Delta t=25$ and the evolution operator $e^{- iH \Delta t}$ was calculated using the function MatrixExp in Mathematica.

%%%%%%%%%%%%%%%%%%%%%%%%%%%%%%%%%%%%%%%%%%%%%%%%%%%%%%%%%%%%%%%%%%%%
\begin{figure}
	\centering
	\includegraphics[width=0.9\linewidth]{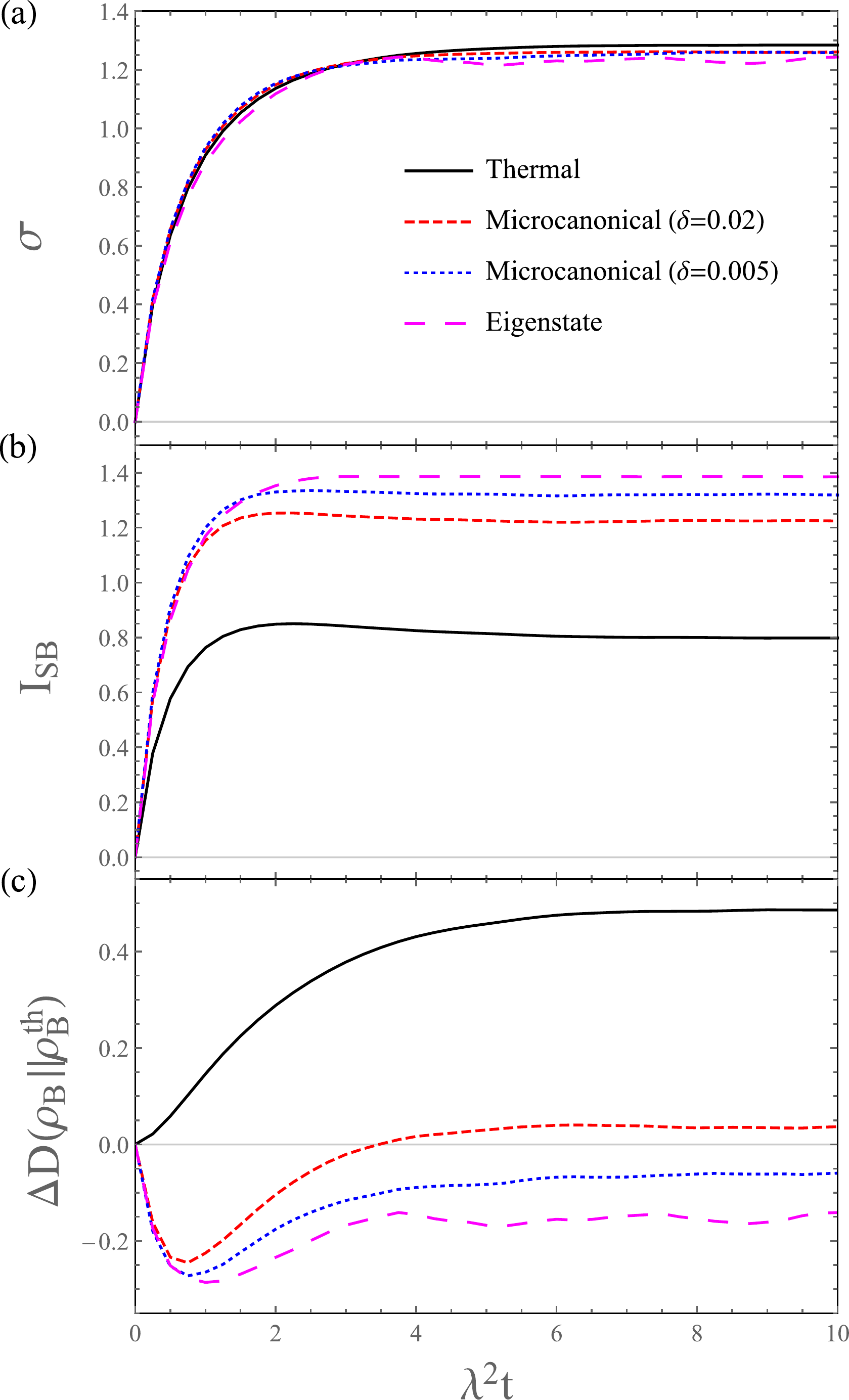}
	\caption{Entropy production $\sigma$ (a) and its information-theoretic constituents $I_{SB}$ (b) and $\Delta D(\rho_B||\rho_B^\text{th})$ (c) as a function of time for the random matrix system with different initial states of the bath. Results for the initial excited state of the system, $\beta=10$, $\gamma=\lambda=0.1$, and $N=2000$.}
	\label{fig:rand}
\end{figure}
%%%%%%%%%%%%%%%%%%%%%%%%%%%%%%%%%%%%%%%%%%%%%%%%%%%%%%%%%%%%%%%%%%%%

The results are presented in Fig.~\ref{fig:rand}. As one can observe, all initial states of the bath generate approximately the same evolution of the entropy production $\sigma$, which confirms their dynamical equivalence. However, the information-theoretic constituents of the entropy production are not the same for different ensembles. In particular, $I_{SB}$ is smallest for the initial canonical state, intermediate for microcanonical states and largest for the pure state; furthermore, its value depends also on the width of the microcanonical shell -- it is larger for smaller widths. We note here a similarity to the previous study of a pure dephasing, showing that different types of the global system-environment dynamics may lead to the same reduced dynamics of the system but generate different system-bath correlations~\cite{smirne2021}. Accordingly, also the relative entropy term $\Delta D(\rho_B||\rho_B^\text{th})$ is different for various initial states. Interestingly, it may also become negative, which implies that the environment is brought closer to the canonical thermal state during the thermalization process; as a matter of fact, when both the system and the bath are initialized in pure states, then $I_{SB}=2\Delta S_S$, and thus $\Delta D(\rho_B||\rho_B^\text{th})$ has to be negative for $-\beta Q< \Delta S_S$. A qualitative interpretation of this observation is well illustrated by the next model.

\textit{Example 2: Noninteracting resonant level}. As a second case, we considered a system with the Hamiltonian of the environment which can be decomposed into a sum of independent modes ($H_B=\sum_k H_k$), such that Eq.~\eqref{entrprodmodgen} is applicable. Specifically, we focused on the noninteracting resonant level model with $H_S = \epsilon_d c_d^\dagger c_d$, $H_B =\sum_{k=1}^K \epsilon_k c_k^\dagger c_k$, and $H_I = \Omega \sum_{k=1}^K \left(c_d^\dagger c_k + c_k^\dagger c_d \right)$, where $c_i^\dagger$ ($c_i$) are creation (annihilation) operators, $\epsilon_i$ are level energies and $\Omega$ is the tunnel coupling. The levels of the environment have been taken to be evenly distributed throughout the interval $[-\Lambda/2,\Lambda/2]$, while the tunnel coupling has been parameterized as $\Gamma={2 \pi \Omega^2 (K-1)/\Lambda}$, where $\Gamma$ is the coupling strength. Parameters have been set as $\Lambda=3 \Gamma$, $\epsilon_d=-0.5 \Gamma$, and $K=7$. he matrix form of the Hamiltonian was obtained using the Jordan-Wigner transformation of the creation and annihilation operators into spin operators: $c_k^\dagger=[\bigotimes_{i=0}^{k-1} (-\sigma_z)] \otimes \sigma_+ \otimes [\bigotimes_{i=k+1}^{K} \mathds{1}_{2}]$ and $c_k=[\bigotimes_{i=0}^{k-1} (-\sigma_z)] \otimes \sigma_- \otimes [\bigotimes_{i=k+1}^{K} \mathds{1}_{2}]$, where $\sigma_\pm =(\sigma_x \pm i \sigma_y )/2$, and $\mathds{1}_{2}$ is $2 \times 2$ identity matrix. The state $\rho_{SB}(t)$ was propagated using the same method as before, with the time step $\Delta t=0.05 \Gamma^{-1}$.

We considered three types of initial states. The first is the microcanonical state, here defined as an equally-weighted mixture of 12 eigenstates of the bath with energy $E=-1.5 \Gamma$ (more precisely, since we fix the energy while enabling the particle number to vary, this may be rather referred to as the grand microcanonical~\cite{lecar1981} or Maxwell's demon ensemble~\cite{navez1997}). Since the system is integrable, and thus ETH is not applicable, the initial pure state is here defined as a superposition of all states from the microcanonical shell: $|\Psi \rangle=W^{-1/2} \sum_i |\psi_i \rangle$; such states provide the dynamical equivalence due to canonical typicality~\cite{tasaki1998, goldstein2006, camalet2008}. Here we took the same amplitude $W^{-1/2}$ for all eigenstates; as shown in the Supplemental Material~\cite{supp}, a good convergence is also observed for randomly chosen amplitudes. Finally, as a third initial state we took the canonical state (or rather, the grand canonical state with the chemical potential $\mu=0$) with the temperature $\beta \approx 0.969 \Gamma$ given by the condition $\text{Tr} (H_B \rho_B^\text{th})=E=-1.5 \Gamma$.

%%%%%%%%%%%%%%%%%%%%%%%%%%%%%%%%%%%%%%%%%%%%%%%%%%%%%%%%%%%%%%%%%%%%
\begin{figure}
	\centering
\includegraphics[width=0.9\linewidth]{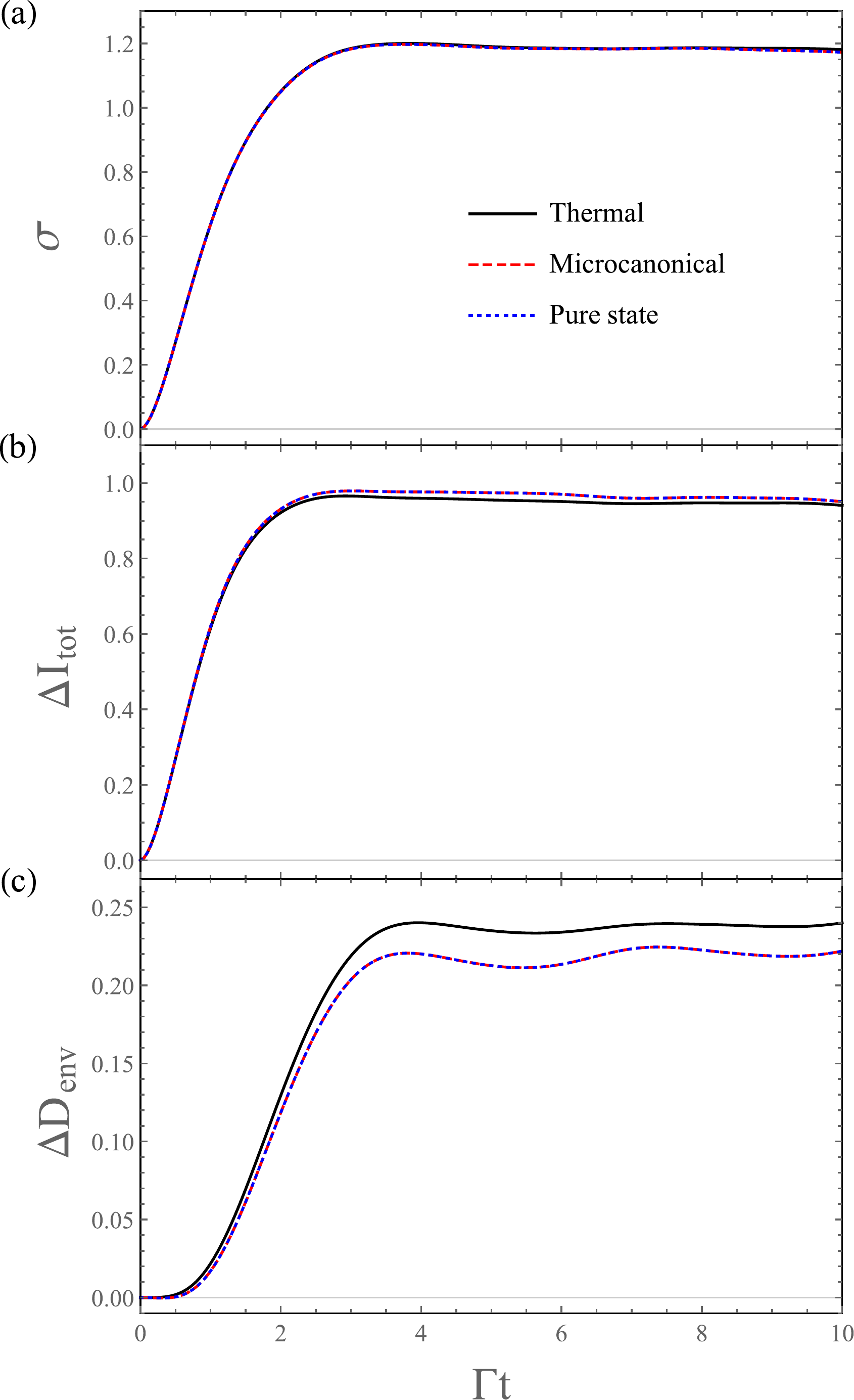}
	\caption{Entropy production (a), change of the total correlation $\Delta I_\text{tot}$ (b), and the change of mode perturbation from equilibrium $\Delta D_\text{env}$ (c) as a function of time for the noninteracting resonant level with different initial states of the environment: the canonical state (black solid line), the microcanonical state (red dashed line), and the pure state (blue dotted line).}
	\label{fig:entr}
\end{figure}
%%%%%%%%%%%%%%%%%%%%%%%%%%%%%%%%%%%%%%%%%%%%%%%%%%%%%%%%%%%%%%%%%%%%

%%%%%%%%%%%%%%%%%%%%%%%%%%%%%%%%%%%%%%%%%%%%%%%%%%%%%%%%%%%%%%%%%%%%
\begin{figure}
	\centering
	\includegraphics[width=0.9\linewidth]{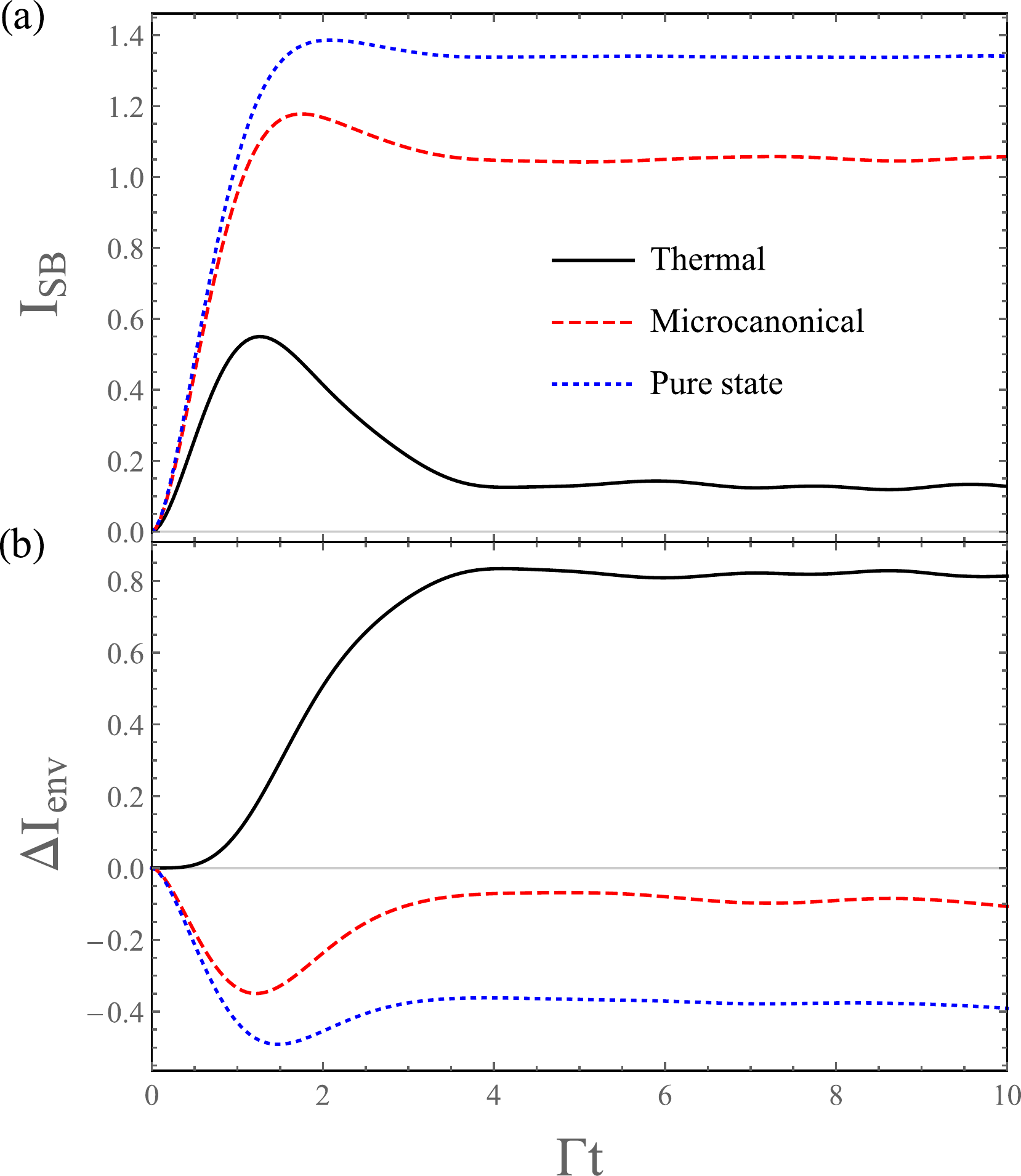}
	\caption{System-bath (a) and intrabath (b) mutual information as a function of time for the noninteracting resonant level with different initial states of the environment; designations as in Fig.~\ref{fig:entr}.}
	\label{fig:inf}
\end{figure}
%%%%%%%%%%%%%%%%%%%%%%%%%%%%%%%%%%%%%%%%%%%%%%%%%%%%%%%%%%%%%%%%%%%%

The results are presented in Figs.~\ref{fig:entr} and~\ref{fig:inf}. As shown in Fig.~\ref{fig:entr}, as argued before, not only the entropy production $\sigma$, but also its information-theoretic constituents $\Delta I_\text{tot}$ and $\Delta D_\text{env}$, are approximately similar for all initial states (to be more precise, the analyzed quantities are exactly the same for microcanonical and pure states; this is because they are fully determined by two-point correlations $\langle c_i^\dagger c_j \rangle$, which for quadratic Hamiltonians evolve independently from higher-order correlations~\cite{peschel2003}, and thus are the same in both cases). Furthermore, $\Delta I_\text{tot}$ is the dominant contribution to the production of entropy, as the perturbation of the modes is relatively small (though there is still a non-vanishing contribution $\Delta D_\text{env}$ due to the finite size of the bath). Therefore, there exists a partial ensemble equivalence regarding the microscopic nature of the entropy production: for all initial states it is mainly related to the change of total correlations between the system and the modes of environment. However, as shown in Fig.~\ref{fig:inf}, the decomposition of $\Delta I_\text{tot}$ into the system-bath and intrabath correlations is not equivalent for different initial states. For the microcanonical and pure states $\Delta I_\text{env}$ is negative which implies that intrabath correlations are destroyed rather than created. This provides a qualitative interpretation for the negativity of $\Delta D(\rho_B||\rho_B^\text{th})$, which has been observed also for the random matrix model: it is related to the destruction of initial non-thermal correlations in the bath.

\textit{Conclusions}. In summary, we have shown that different initial states of the bath leading to the same reduced dynamics and thermodynamics on the system may be only partially equivalent with respect to the information-theoretic formulation of the entropy production: while in all cases the entropy production can be expressed as a sum of two information-theoretic contributions $I_{SB}$ and $\Delta D(\rho_B||\rho_B^\text{th})$, the relative weight of those terms may depend on the initial state. A particular instance of equivalence of the information-theoretic contributions can be observed for environments composed of independent modes: for all ensembles of the bath the entropy production is mostly related to the change of total correlation between the system and the modes of environment $\Delta I_\text{tot}$. This supports a general idea relating the entropy production to the generation of multipartite correlations~\cite{esposito2010, ptaszynski2019}. However, again, the decomposition of $\Delta I_\text{tot}$ into the change of system-bath and intrabath correlations may vary for different initial states of the environment.

It is important to note that the difference between various initial states can be significant only when the contributions $I_{SB}$ and $\Delta D(\rho_B||\rho_B^\text{th})$ are of a similar order of magnitude. As discussed by us previously~\cite{ptaszynski2019}, this can be only true when the entropy production is sufficiently small, since the system-bath mutual information is bounded from above by the Araki-Lieb inequality $I_{SB} \leq 2 \ln \text{dim} \mathcal{H}_S$~\cite{araki1970}, where $\text{dim} \mathcal{H}_S$ is the dimension of the Hilbert space of the system. As a result, for $\sigma$ significantly exceeding $2 \ln \text{dim} \mathcal{H}_S$ the entropy production becomes dominated by the displacement term [$\sigma \approx \Delta D(\rho_B||\rho_B^\text{th})$], independently of the initial state of the bath. Thus, dynamically equivalent states of the bath become also informationally equivalent in the limit of large entropy production. This resembles the standard concept of ensemble equivalence which states that different ensembles become equivalent in the thermodynamic limit of a large system.

In addition to thermodynamics, the present study may be relevant for the field of condensed matter physics, where correlations between quantum impurities and their environment have gained a certain amount of attention~\cite{yoo2018, kim2021}. Our results suggest that such correlations may depend on the choice of the ensemble. In particular, beyond the cases analyzed in our paper, it is worth investigating whether there is a difference between the grand-canonical ensemble and the canonical ensemble with a fixed particle number, which may be more physically justified, e.g., in the description of impurities coupled to trapped ultracold atoms~\cite{bauer2013}.

\begin{acknowledgments}
K. P. has been supported by the National Science Centre, Poland, under Project No. 2017/27/N/ST3/01604, and by the Scholarships of Minister of Science and Higher Education. This research was also supported by the FQXi foundation project FQXi-IAF19-05.
\end{acknowledgments}

\end{document}

%\preprint{APS/123-QED}
	
\title{Supplemental Material to: Ensemble dependence of information-theoretic contributions to the entropy production}% Force line breaks with \\
	
\author{Krzysztof Ptaszy\'{n}ski}
\affiliation{Complex Systems and Statistical Mechanics, Department of Physics and Materials Science, University of Luxembourg, L-1511 Luxembourg, Luxembourg}
\affiliation{Institute of Molecular Physics, Polish Academy of Sciences, Mariana Smoluchowskiego 17, 60-179 Pozna\'{n}, Poland}
\email{krzysztof.ptaszynski@uni.lu}
\author{Massimiliano Esposito}
\affiliation{Complex Systems and Statistical Mechanics, Department of Physics and Materials Science, University of Luxembourg, L-1511 Luxembourg, Luxembourg}

\date{\today}% It is always \today, today,

\maketitle

\subsection*{Noninteracting resonant level model with random pure initial states of the bath}

%
%%%%%%%%%%%%%%%%%%%%%%%%%%%%%%%%%%%%%%%%%%%%%%%%%%%%%%%%%%%%
\begin{figure}
	\centering
	\includegraphics[width=0.8\linewidth]{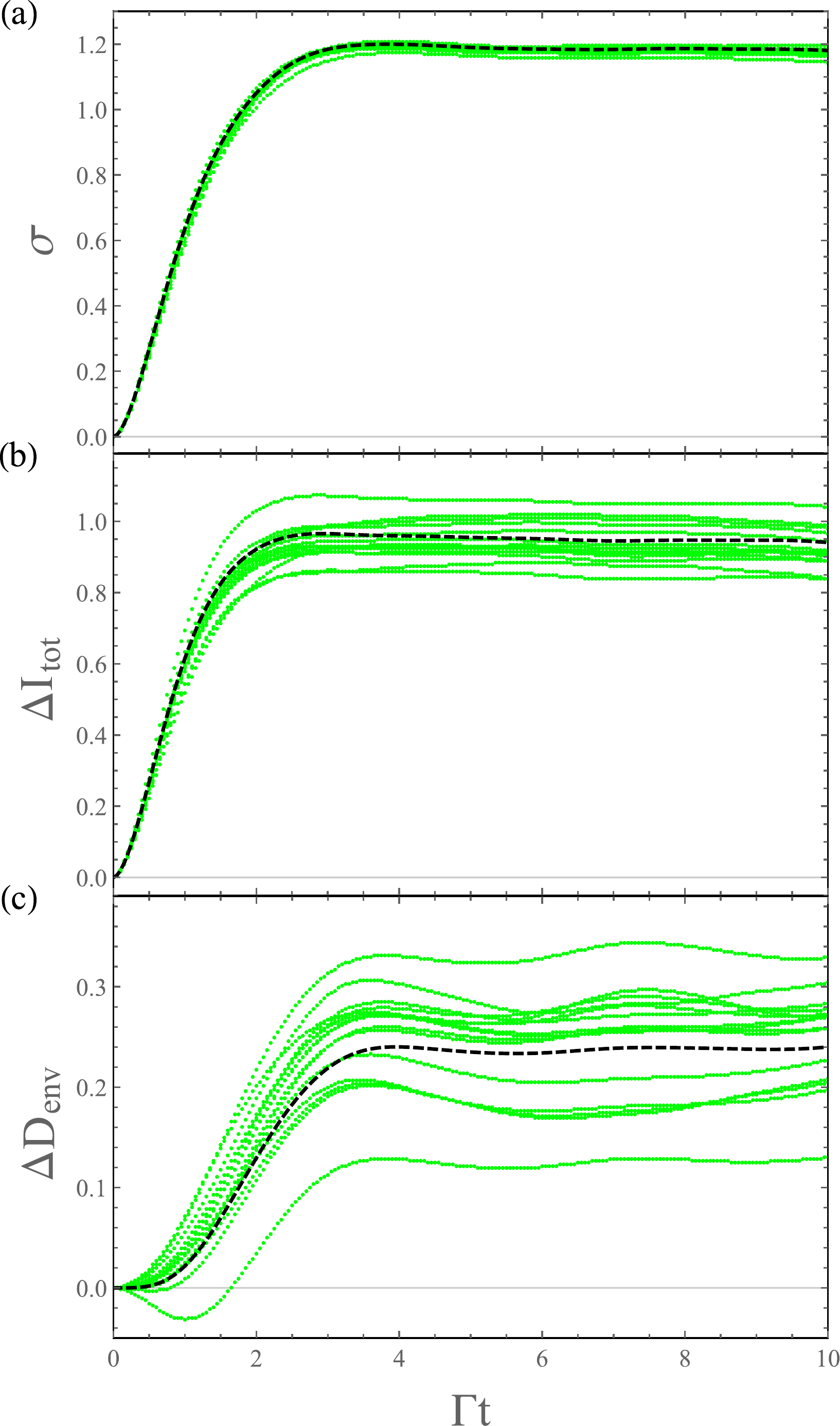}		
	\caption{Entropy production and its constituents $\Delta I_\text{tot}$ and $\Delta D_\text{env}$ for 15 random initial pure states of the bath (green solid lines) compared with the trajectories obtained for the thermal state (black dashed line). Parameters as in the main text.}
	\label{fig:randpurj1}
\end{figure}
%%%%%%%%%%%%%%%%%%%%%%%%%%%%%%%%%%%%%%%%%%%%%%%%%%%%%%%%%%%%
%
%
%%%%%%%%%%%%%%%%%%%%%%%%%%%%%%%%%%%%%%%%%%%%%%%%%%%%%%%%%%%%
\begin{figure}
	\centering
	\includegraphics[width=0.8\linewidth]{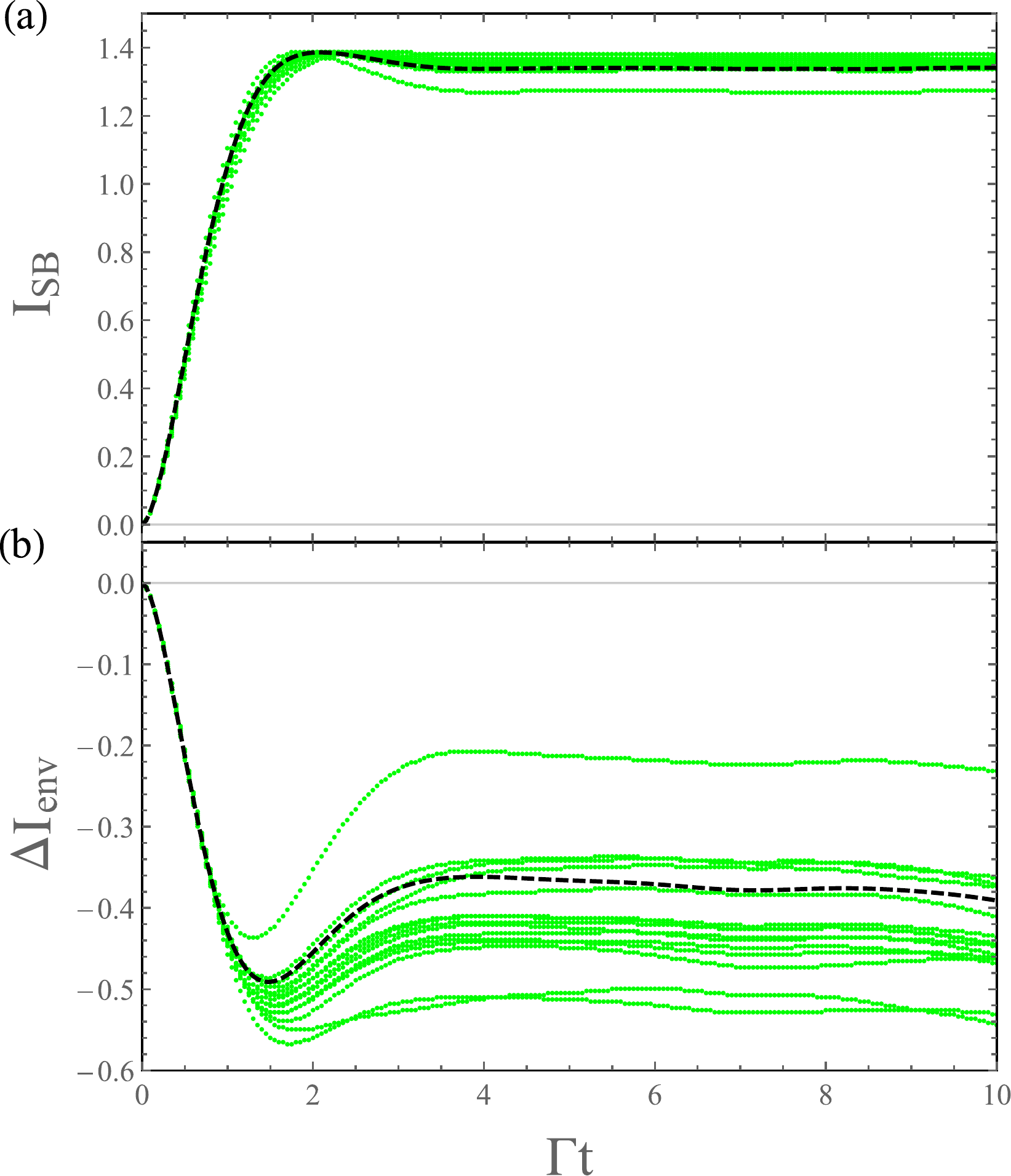}		
	\caption{The system-bath mutual information $I_{SB}$ and the change of intraenvironment correlation $\Delta I_\text{env}$ for 15 random initial pure states of the bath (green solid lines) compared with the trajectories obtained for the equally-weighted superpositions of eigenstates from the microcanonical shell (black dashed line). Parameters as in the main text.}
	\label{fig:randpurj2}
\end{figure}
%%%%%%%%%%%%%%%%%%%%%%%%%%%%%%%%%%%%%%%%%%%%%%%%%%%%%%%%%%%%
%

In the main text we investigated the behavior of entropy production and its information-theoretic constituents in the noninteracting resonant level model for a pure initial state of the bath defined as an equally-weighted superposition of eigenstates from the microcanonical shell. However, as implied by the principle of canonical typicality~\cite{goldstein2006}, thermalization should take place (in the thermodynamic limit) also for a typical randomly chosen superposition of eigenstates; therefore, an approximate convergence should be observable also for finite baths. Here we study the evolution of analyzed quantities for 15 random initial states defined as $|\Psi \rangle=\sum_j \alpha_j |\psi_j \rangle$, where $|\psi_j \rangle$ are eigenstates from the microcanonical shell and $\alpha_j$ are random probability amplitudes defined as $\alpha_j=\sqrt{r_j/R} e^{2\pi i \phi_j}$, where $r_j$ and $\phi_j$ are chosen from a uniform distribution over the interval $[0,1]$, and $R=\sum_j r_j$.

The results are presented in Figs.~\ref{fig:randpurj1} and~\ref{fig:randpurj2}. As one can observe, the calculated entropy production and its constituents are spread-out around the results obtained for the thermal state [Fig.~\ref{fig:randpurj1}] or the equally-weighted superposition of eigenstates [Fig.~\ref{fig:randpurj2}]. Interestingly, the obtained trajectories of the entropy production seem to be more concentrated around the thermal value than those of its constituents $\Delta I_\text{tot}$ and $\Delta D_\text{env}$. This shows that the precision of simulating the open system dynamics using pure states of the bath (e.g., within the framework of dynamical typicality~\cite{heitmann2020}) may depend on the analyzed quantity.